# Polaronic hole localization and multiple hole binding of acceptors in oxide wide-gap semiconductors

Stephan Lany and Alex Zunger

*National Renewable Energy Laboratory, Golden, CO 80401*

Acceptor-bound holes in oxides often localize asymmetrically at one out of several equivalent oxygen ligands. Whereas Hartree-Fock (HF) theory overly favors such symmetry-broken polaronic hole-localization in oxides, standard local density (LD) calculations suffer from spurious delocalization among several oxygen sites. These opposite biases originate from the opposite curvatures of the energy as a function of the fractional occupation number $n$, i.e., $d^2E/dn^2 < 0$ in HF and $d^2E/dn^2 > 0$ in LD. We recover the correct linear behavior, $d^2E/dn^2 = 0$, that removes the (de)localization bias by formulating a generalized Koopmans condition. The correct description of oxygen hole-localization reveals that the cation-site nominal single-acceptors in ZnO, $In_2O_3$, and $SnO_2$ can bind multiple holes.

Experimental evidence for cation-site acceptors in wide-gap oxides demonstrates that holes often lock into individual oxygen ligands instead of being distributed over all symmetry-equivalent oxygen sites, e.g., in case of Al$_{Si}$ in SiO$_2$ or Li$_{Zn}$ in ZnO [1, 2]. Formally, the hole binding can be described by the change of the oxidation state of individual O atoms from the normal O$^{-II}$ in to an O$^{-I}$ state [1, 2]. A systematically consistent theoretical description of such acceptor states remains, however, challenging as common density-functional theory (DFT) calculations in the local density or generalized gradient approximations (LDA or GGA) fail to reveal the correct hole localization on just one O-atom and the associated lattice relaxation effects. This failure has been traced back to a residual self-interaction present within the O-$p$ shell in LDA or GGA [3, 4, 5]. Hartree-Fock (HF) theory on the other hand, is known to overestimate the tendency towards hole-localization on individual lattice sites [2]. The correct description of the balance between competing tendencies towards single-site localization *vs.* delocalization among equivalent sites, bears great importance for physical phenomena like hyperfine interactions [4], magnetism without *d*-elements [5, 6], and *p*-type doping in wide-gap oxides [7, 8].

A number of electronic structure methods are able to restore the qualitative picture of single-site localization, e.g., self-interaction correction (SIC) [4, 5], DFT+U [6, 9], and hybrid-DFT [10, 11]. However, a guiding principle as to how to select free parameters (e.g., $U$ in DFT+U or the fraction of HF exchange in hybrid-DFT) has been missing. In order to achieve a systematical quantitative description of hole-localization, we first formulate the condition that assures that the energies of the unoccupied hole states are correctly placed relative to the spectrum of occupied states. Second, we define an on-site occupation-dependent potential that serves to increase, relative to DFT, the energy splitting between occupied and unoccupied states. By determining the strength of this potential such that the general condition of the first step is fulfilled, we achieve a consistent description of localization *vs.* delocalization without relying on empirical parameters. Applying this method to cation-site acceptors in ZnO, In$_2$O$_3$, and SnO$_2$, we find that these impurities can bind multiple holes even though from their position in the periodic table they are expected to be single acceptors, and we demonstrate the importance of the correct description of hole-localization for the prediction of *p*-type doping of oxides.

*Condition for the correct splitting between occupied and unoccupied states.* Due to residual self-interaction [12], the energy splitting between occupied and unoccupied states is generally underestimated in approximate DFT functionals like LDA or GGA. This is particularly true for atomic *d*- and *f*-orbitals, but also for the *p*-orbitals of anionic first-row elements like O or N, which are sometimes described as "strongly correlated" systems [6]. The key for the correct placement of unoccupied states relative to the spectrum of occupied states lies in the observation that LDA/GGA and HF-theory produce errors of opposite sign in the energy *E* between adjacent integer occupation numbers $n_i$ of the highest occupied state *i* [13, 14]: As illustrated schematically in Fig. 1a, in HF-theory the energy is a *concave* function of the (continuous) occupation number, $d^2E/dn_i^2 < 0$, but it is generally a *convex* function, $d^2E/dn_i^2 > 0$, in LDA or GGA. The correct behavior, however, would be linear [13, 14], i.e.,

$$d^2E/dn_i^2 = 0 \qquad (1)$$

On the other hand, the electron addition energy (under fixed atomic positions) can be expressed as [12]

$$E(N+1) - E(N) = e_i(N) + \Pi_i + \Sigma_i, \qquad (2)$$

where $e_i(N)$ is the energy eigenvalue before electron addition, $\Pi_i$ is the self-interaction energy after electron addition to the orbital *i* under the constraint of the wave-functions being fixed at the initial-state, and $\Sigma_i$ is the energy contribution arising due to wave-function relaxation. In the case of the correct linear behavior [eq. (1)], the integration of Janak's theorem, $dE(n_i)/dn_i = e_i(n_i)$, leads to the condition

$$E(N+1) - E(N) = e_i(N), \text{ or} \qquad (3)$$

$$\Pi_i + \Sigma_i = 0.$$

While resembling Koopmans theorem in Hartree-Fock theory, eq. (3) formulates a *condition* to be fulfilled, whereas the original Koopmans theorem states an *approximate equality*. Indeed, in HF theory, where $\Pi_i \equiv 0$, the condition eq. (3) is generally not fulfilled since $\Sigma_i < 0$ is not negligible for localized defect or impurity states in semiconductors. Consequently, the initially unoccupied eigenvalue $e_i$ is lowered upon electron addition in HF (see Fig. 1b). In contrast, in LDA or GGA, where $\Pi_i > 0$ and generally also $\Pi_i + \Sigma_i > 0$, the energy of the eigenvalue $e_i$ increases following the electron addition (Fig. 1b) as a result of residual self-interaction.

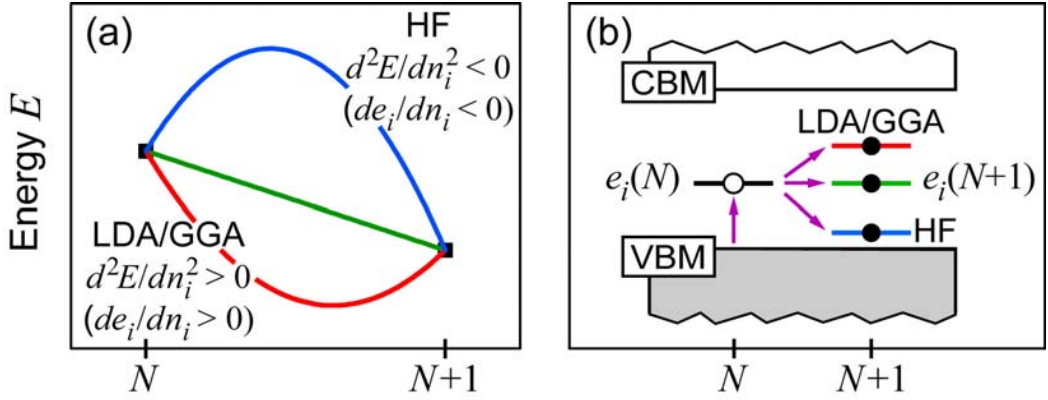

Figure 1 (color online). (a) In HF-theory (blue) the energy is a concave function of the continuous occupation number $n_i$, but a convex one in LDA or GGA (red). The correct behavior (green) is linear (see Refs. [13, 14]). (b) The corresponding shift of the eigenvalue $e_i$ upon electron addition into a localized hole state in an oxide (the vertical arrow indicates optical excitation from the VBM into the hole state).

*How to make the generalized Koopmans condition satisfied.* Due to the opposite behavior of LDA/GGA and HF (Fig. 1), methods that introduce HF-like interactions into the DFT Hamiltonian, such as hybrid-DFT or LDA+U (when applied to O-*p*) are conceptually justified to correct the spurious delocalization, and their parameters could in principle be adjusted to make eq. (3) satisfied. However, the parameters that are suited to remove the shift of the energy of the hole state upon occupation (Fig. 1b) are in general not simultaneously appropriate for the description of the host matrix. For example the fraction of 42% HF (exact) exchange used in Ref. [10] for the case of $SiO_2$:Al largely exceeds the fraction of 25% that is generally more suitable for the band structure of wide gap semiconductors [15]. A similar concern exists for DFT+U, which creates an additional potential of the form [16]

$$V_U = (U - J)(0.5 - n_{m,\sigma}),  \qquad (4)$$

where, $n_{m,\sigma}$ is the fractional occupancy (partial charge) $0 \leq n_{m,\sigma} \leq 1$ of the *m*-sublevel of spin $\sigma$ (diagonal representation). Since, the occupancy of O-*p* orbitals in the oxides considered here depends strongly on the projection radius [17], the application of DFT+U to O-*p* orbitals distorts the band-structure of the defect-free oxide host in a rather uncontrolled way.

In order to satisfy the generalized Koopmans condition, eq. (3), without any adverse side-effects, we define a potential operator that acts exclusively on the empty hole state (hs),

$$V_{\text{hs}} = \lambda_{\text{hs}}\left(1 - n_{m,\sigma}/n_{\text{host}}\right). \tag{5}$$

The quantity $n_{\text{host}}$ is the occupancy of the respective $lm$-channel in the unperturbed host (e.g., O-$p$), and $\lambda_{\text{hs}}$ is a parameter for the strength of the potential that is determined through the condition eq. (3). The technical implementation of the potential operator $V_{\text{hs}}$ is achieved through a combination of the occupation dependent DFT+U potential, eq. (4), and our (occupation-independent) non-local external potential $V_{\text{nlep}}$ of Ref. [18]. The hole-state potential $V_{\text{hs}}$, eq. (5), retains the conceptual justification of DFT+U, but eliminates, by construction, the effect on the host band structure. Thus, applying $V_{\text{hs}}$ to the O-$p$ orbitals in an otherwise standard GGA calculation allows us to stabilize the localized hole-states for acceptor-bound polarons [1, 2], and to determine their correct energies relative to the spectrum of occupied states as determined in GGA. We emphasize that the value of the parameter $\lambda_{\text{hs}}$ is defined through the general condition eq. (3) and does not rely on any empirical data. Therefore, the *first-principles* character of the GGA Hamiltonian is not compromised. All present calculations are performed using the projector augmented wave (PAW) implementation of the VASP code [19], and are based on the GGA of Ref. [20]. Supercell size effects have been treated as described in Ref. [21], where, in particular, we showed that finite-size effects for charged defects are effectively eliminated when the 3$^{\text{rd}}$ order image charge interaction and potential alignment effects are taken into account simultaneously.

*Application of $V_{hs}$ to ZnO:Li.* Figure 2a shows the structural and magnetic properties around the Li$_{\text{Zn}}$ impurity in ZnO as a function of $\lambda_{\text{hs}}$. Whereas in a standard LDA or GGA calculation the wavefunction of the unoccupied state has a practically equal amplitude at all four O neighbors and decays only very slowly with the distance from the Li$_{\text{Zn}}$ site, the hole locks into a single O-$p$ orbital, as shown in Fig. 3a, when $\lambda_{\text{hs}}$ exceeds a critical (cr) value $\lambda_{\text{hs}}^{cr} \approx 3$ eV. As seen in Fig. 2a, this transition entails the spontaneous breaking of the approximate tetrahedral symmetry in GGA and the emergence of a local magnetic moment at the O$^{-1}$ ion that traps the hole. Figure 2b shows the energy eigenvalue $e_i(N)$ of the initially unoccupied state (*cf.* Fig. 1b) and the electron addition energy $E_{\text{add}} = E(N+1) - E(N)$ as a function of $\lambda_{\text{hs}}$. We see that the condition eq. (3) is fulfilled for $\lambda_{\text{hs}}^{lin} = 4.3$ eV, at which point the correct linear (lin) behavior is recovered [see eq. (1) and Fig. 1]. Since, $\lambda_{\text{hs}}^{lin} > \lambda_{\text{hs}}^{cr}$ lies well above the critical value required to stabilize the polaronic state (see Fig. 2a), this symmetry-broken localized state is predicted to be the physically correct state, in agreement with experiment [1]. We further confirmed that our method correctly predicts a delocalized hole state in cases where this is the physical reality, e.g., in the defect free ZnO host (i.e., no hole self-trapping is predicted), and in case of the shallow Li acceptor in ZnTe.

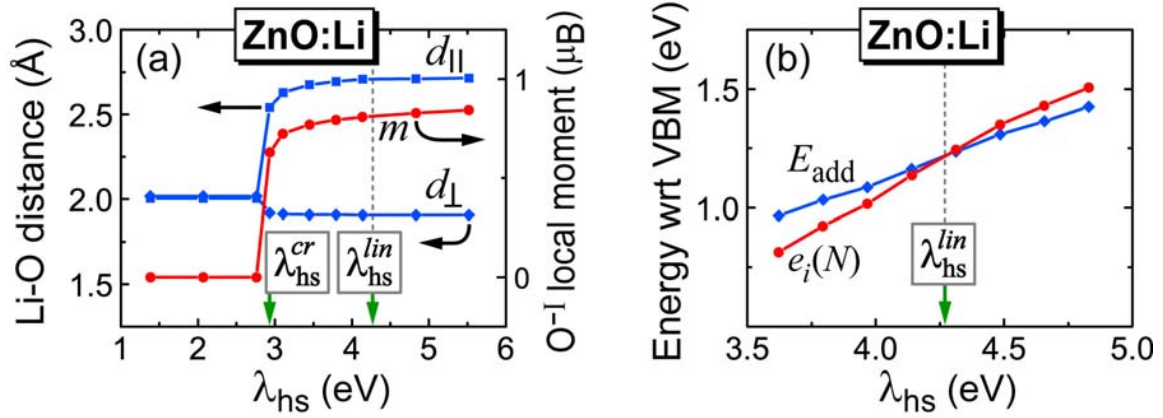

Figure 2 (color online). (a) Structural and magnetic properties of the $Li_{Zn}$ impurity in ZnO, as a function of the hole-state potential strength $\lambda_{hs}$. The polaronic state is stable above a critical value $\lambda_{hs} > \lambda_{hs}^{cr}$. $d_\parallel$: Li-O$^{-I}$ distance; $d_\perp$: Li-O$^{-II}$ distance (cf. Fig. 3a); $m$: local magnetic moment of O$^{-I}$ (integration radius $R = 1$ Å). (b) The electron addition energy $E_{add} = E(N+1)-E(N)$ and the energy eigenvalue $e_i(N)$ of the initially unoccupied acceptor state of Li. $\lambda_{hs}^{lin}$ marks the value of $\lambda_{hs}$ for which eq. (3) is satisfied.

As seen in Fig. 2b, the electron addition energy increases continuously with the strength $\lambda_{hs}$ of the hole-state potential in the vicinity of the correct strength $\lambda_{hs}^{lin}$, whereas the structural properties and the local magnetic moment at the O$^{-I}$ ion are rather insensitive to the value of $\lambda_{hs}$, once the threshold $\lambda_{hs}^{cr} \approx$ 3 eV, (Fig. 2a) for the stabilization of the polaronic state is exceeded. Thus, properties like the lattice distortion, hyperfine parameters [4], or magnetic interactions [5, 6] can generally be expected to be relatively insensitive to the question how well the condition eq. (3) is fulfilled, if only the applied method stabilizes the polaronic state. In contrast, the accurate prediction of quantities which are based on the energy difference between occupied and unoccupied states, e.g., the acceptor binding energies that are addressed next, requires that the fundamental condition [eq. (3)] for the energy splitting between occupied and unoccupied states be fulfilled rather accurately.

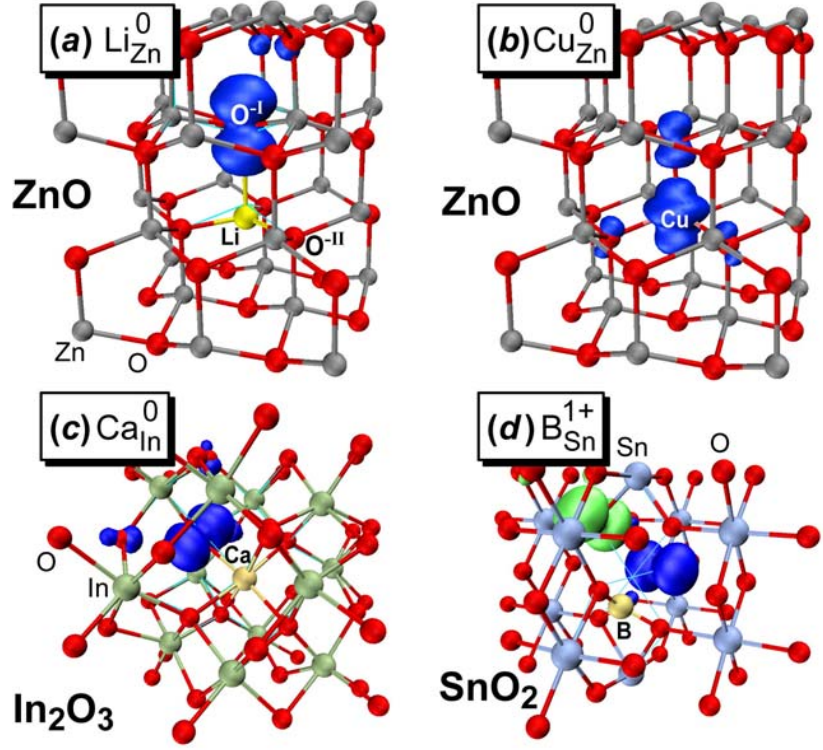

Figure 3 (color online). The calculated atomic structures and the one-particle hole-densities for different cation-site acceptors in ZnO, In$_2$O$_3$, and SnO$_2$ (isosurface-density: 0.03 $e$/Å$^3$). For B$_{Sn}^+$ in SnO$_2$, binding two holes, the low-spin singlet state ($S = 0$; blue = spin-down, green = spin-up) is shown, so to illustrate the localization of the individual one-particle densities. The underlying wire-grid (light blue) shows the host lattice before atomic relaxation.

*Acceptor ionization energies in ZnO, In$_2$O$_3$ and SnO$_2$.* When we apply the hole state potential to an otherwise standard GGA calculation, we obtain the acceptor transition of Li$_{Zn}$ at $\varepsilon(0/-) = E_V+0.36$ eV above the valence band maximum (VBM), considerably deeper than in an uncorrected GGA calculation with a delocalized acceptor state ($E_V+0.16$ eV). While the hole state operator, eq. (5), allows to correctly place the energy of the hole state relative to the occupied host bands in a non-empirical way, it does not correct the strong band-gap underestimation of GGA ($E_g = 0.7$ eV). In order to avoid an ambiguity of the position of the acceptor level inside the true band-gap, we present in Table 1 results for metal-site acceptors in ZnO, In$_2$O$_3$, and SnO$_2$, that are obtained with an additional empirical band-gap correction via the non-local external potentials [22]. The $\lambda_{hs}$ parameters for the band-gap corrected calculations are determined via eq. (3) as 4.8, 4.3, and 4.1 eV for ZnO, In$_2$O$_3$, and SnO$_2$, respectively.

Table 1. The calculated acceptor ($\varepsilon_A$) and ultradeep donor ($\varepsilon_D$) levels of nominal single-acceptors in ZnO, In$_2$O$_3$, and SnO$_2$ (in eV relative to the VBM). For Cu and Ag, the values in parenthesis include an additional hole-state correction for their *d*-states.

| ZnO | Li | Na | Cu | Ag |
|---|---|---|---|---|
| $\varepsilon_A(0/-) = E_V +$ | 0.86 | 0.79 | 2.32 (3.46) | 1.18 (1.54) |
| $\varepsilon_D(+/0) = E_V +$ | 0.15 | 0.25 | 0.37 (0.14) | 0.56 (0.35) |
| **In$_2$O$_3$** | **Be** | **Mg** | **Ca** | **Zn** |
| $\varepsilon_A(0/-) = E_V +$ | 1.27 | 0.86 | 0.68 | 1.04 |
| $\varepsilon_D(+/0) = E_V +$ | 0.97 | 0.41 | 0.37 | 0.63 |
| **SnO$_2$** | **B** | **Al** | **Ga** | **In** |
| $\varepsilon_A(0/-) = E_V +$ | 1.57 | 0.85 | 0.76 | 0.58 |
| $\varepsilon_D(+/0) = E_V +$ | 1.39 | 0.47 | 0.44 | 0.30 |

We see in Table 1 that the acceptor level of Li at 0.9 eV above the VBM is now closer to the experimental value of 0.8 eV [23] compared to the calculation without the band-gap correction. The latter underestimates the acceptor binding energy mainly because of overestimated *p-d* repulsion between the valence band states and the Zn-*d* shell, placing the VBM too high [21] relative to the Li acceptor state. In the band-gap corrected calculation, the Zn-*d* energies are corrected by GGA+U [22]. We find that all acceptor states are localized in a single O-*p* orbital (see Fig. 3), and that the ionization energies are relatively deep, $\varepsilon_A \geq 0.6$ eV (Table 1), thereby strongly questioning the suitability of cation-site acceptors in ZnO, In$_2$O$_3$, and SnO$_2$ for the purpose to achieve *p*-type transparent conductive oxides [7, 8]. In addition to the $\varepsilon(0/-)$ acceptor level, we observe also an ultra-deep $\varepsilon(+/0)$ donor level closer to the VBM, which results from the binding of a second hole. When two holes are bound in the positive charge state, the two individual one-particle states are each localized on only a single O-*p* orbital, as shown for $B_{Sn}^+$ in SnO$_2$ (Fig. 3d), similar like in case of the charge-neutral Zn vacancy $V_{Zn}$ in ZnO (Fig. 3b). In case of the strongly size-mismatched impurities Be$_{In}$ in In$_2$O$_3$ and for B$_{Sn}$ in SnO$_2$, we further observe a large lattice relaxation that breaks three Be-O or B-O bonds and leads to threefold coordination (Fig. 3d), even in the ionized $Be_{In}^-$ and $B_{Sn}^-$ states. The breaking of three bonds with O-neighbors facilitates the binding of yet another hole leading to a deep (double) donor transition at $\varepsilon_D(2+/+) = E_V + 0.21$ eV for Be$_{In}$ in In$_2$O$_3$ and at $E_V + 0.62$ eV for B$_{Sn}$ in SnO$_2$.

*Group Ib acceptors in ZnO.* In case of the group Ib elements Cu and Ag, the hole-wavefunction has mainly the *d*-character of the impurity atom (see Fig. 3b), and not O-*p* character like in case of the main group acceptors. Thus, the hole-state potential for O-*p* has little effect and is not sufficient to satisfy the condition eq. (3). Therefore, we apply GGA+U with $(U-J) = 5$ and 4 eV for Cu-*d* and for Ag-*d*, respectively [22]. These parameters have been established before, and were chosen so to reproduce photoemission spectra, e.g., for $Cu_2O$, i.e., to correctly place the energies of the occupied *d*-shell electrons [21]. Despite the application of GGA+U, the condition eq. (3) is not fulfilled satisfactorily yet, which indicates that the energy splitting between occupied and unoccupied *d*-symmetries is still underestimated [24]. Therefore, we apply an additional hole-state potential $V_{hs}$ on Cu-*d* and Ag-*d* so that condition eq. (3) is satisfied. In Table 1, we give the transition energies for $Cu_{Zn}$ and $Ag_{Zn}$ with and without this additional correction. In either case, we find very deep acceptor states which are not conducive to *p*-type doping, in contrast to much more optimistic conclusion for Ag derived from standard LDA calculations [7]. Like all other acceptors, Cu and Ag can bind a second hole, giving rise to a deep donor transition (Table 1). In contrast to the other cases, however, where both holes are bound in a O-*p* orbital (Fig. 3d), here only one hole is located in a O-*p* orbital while the other is in the Cu- or Ag-*d* shell, e.g., forming a $Cu^{+II}(d^9)+O^{-I}$ configuration.

*Hole-binding of $V_{Zn}$ in ZnO.* Previous DFT calculations of $V_{Zn}$ predicted acceptor levels close above the VBM, which, however, could not be reconciled with magnetic resonance data [25]. Nevertheless, as we have shown in Ref. [26], the O-*p* dangling bonds lie inside the band gap in a standard GGA calculation, which suggests that $V_{Zn}$ could bind up to 4 holes, one at each O neighbor, if the delocalization due to the self-interaction error is avoided. Indeed, applying the present method to $V_{Zn}$ we find that all charge states from +2, ..., –2 lie within the gap, with the respective transition levels being located at 0.45, 0.99, 1.46, and 1.91 eV above the VBM. Notably, the binding of up to 4 holes at $V_{Zn}$ is also supported by a recent hybrid-DFT calculation [11].

*Conclusions.* In order to place in a DFT calculation the O-*p* bound hole-states of acceptors in oxides at their correct energies relative to the spectrum of occupied states, we introduced a hole-state potential operator that that acts only on the hole states, but does not affect the underlying host band-structure.

Quantitative predictions of acceptor binding energies were attained by non-empirical determination of the parameter for the potential strength through a generalized Koopmans condition.

This work was funded by the U.S. Department of Energy, Office of Energy Efficiency and Renewable Energy, under Contract No. DE-AC36-08GO28308 to NREL. The use of MPP capabilities at the National Energy Research Scientific Computing Center is gratefully acknowledged.